\documentclass[hyperref,twocolumn,pra]{revtex4}%
\usepackage{amsfonts}
\usepackage{amsmath}
\usepackage{amssymb}
\usepackage{graphicx}%
\providecommand{\U}[1]{\protect\rule{.1in}{.1in}}

\begin{document}
\title[Analytic Results for NLCS]{Exact final state integrals for strong field QED}
\author{Victor Dinu}
\affiliation{Department of Physics, University of Bucharest, P. O. Box MG-11, M\u{a}gurele
077125, Romania}
\pacs{PACS number}

\begin{abstract}
This paper introduces the exact, analytic integration of all final state
variables for the process of nonlinear Compton scattering in an intense plane
wave laser pulse, improving upon a previously slow and challenging numerical
approach. Computationally simple and insightful formulae are derived for the
total scattering probability and mean energy-momentum of the emitted
radiation. The general form of the effective mass appears explicitely. We
consider several limiting cases, and present a quantum correction to Larmor's
formula. Numerical results are plotted and analysed in detail.

\end{abstract}
\startpage{1}
\endpage{2}
\maketitle

\paragraph{Introduction:}

The recent and predicted progress in laser technology leading to very high
peak intensities justify the need for a better understanding of so-called
nonlinear QED, describing phenomena occurring in fields so strong that their
effects cannot be treated perturbatively.

Unfortunately, the complexity of the processes inside these ultra-intense
laser beams has meant that several simplifications have had to be used to make
practical computations feasible. The laser beam is usually supposed to be in a
coherent state, which can be well approximated by a classical field. Due to
the relatively small frequencies, unless massive particles of very high
Lorentz factor are involved, quantum effects are small, so a fully classical
description may often be justified. For instance, in discussing the scattering
of an electron in a laser beam, one may consider Thompson scattering instead
of its quantum counterpart, nonlinear Compton scattering (NLCS) \cite{piata}.
The classical approximation allows for a realistic description of the laser
field and the inclusion of radiation reaction (RR) \cite{diprr}, but is unable
to describe important quantum effects, such as nonperturbative pair creation
from vacuum \cite{pere}, the trident process \cite{trident}, or vacuum
birefringence \cite{biref}.

A treatment of these processes in the framework of nonlinear QED, even in a
semiclassical approach, has not yet been performed without further
approximations, such as replacing the laser field by an idealized plane wave,
thus allowing for analytical (Volkov) solutions to the Dirac equation. This
disregards the strong spatial focusing of the beam needed to attain high
intensities. In addition, for a long time, results were restricted to
infinite, monochromatic plane waves, or even, giving up periodicity, to a
crossed field model \cite{niki1, niki2}. Only recently, the more realistic
short pulse plane waves came into use \cite{boca1, SHK, SK}.

In \cite{boca1, MDP, boca3}, the photonic and electronic distributions
resulting from NLCS were described in detail for some model pulses. In
principle, by integrating these distributions, the total probability and
expectation values, such as for the emitted radiation's energy-momentum, can
be obtained. However, previous papers did never plot these quantities, because
of the great numerical challenge posed by this task. By a change in
integration order and a different regularization, allowing for all final state
integrals to be performed analytically, we obtain formulae that are not only
very easy to compute, but offer a new understanding of the scattering
mechanism, the effective mass's role, the time coherence of the process and
its relation to the classicality parameter. An extensive numerical exploration
of the results, their dependence on the various parameters, and their limits,
now becomes possible. The method we develop is quite general and shall be
applied to other strong field QED processes in a future paper.

\paragraph{Preliminaries:}

For our starting point, notations and conventions, we refer the reader to
\cite{unip}. In short, $p=mv$ and $k^{\prime}$ are the initial electron and
final photon four-momenta. We opt for natural units, so $c=\hbar=1$. Define
$k=\omega n$, where $n^{2}=0$ and $\omega$ is some characteristic frequency of
the wave. Let $\phi=k\cdot x=\omega x^{-}$ be an invariant \textit{lightfront}
coordinate, used to describe the plane wave pulse by the four-potential
$A=\frac{m}{e}a_{0}f\left(  \phi\right)  $. By transversality, $k\cdot A=0$.
We choose $n=\left(  1,\mathbf{e}_{3}\right)  $, $f_{0}=f_{3}=0$, and use
lightfront notations such as $p^{\pm}=p^{0}\pm p^{3}$ $,\ \mathbf{p}^{\perp
}=\mathbf{p}-p^{3}\mathbf{n}$. The final results for probability/momentum will
prove to be manifestly Lorentz invariant/covariant. While working with
$f\left(  \phi\right)  $, only the gauge changes keeping $A$ $\phi$-only
dependent are allowed. But the end results can be expressed in terms of the
classical velocity, so they are gauge invariant. For a long pulse, one may
choose $\omega$ the carrier frequency and the peak value of the envelope of
$f^{\prime}\left(  \phi\right)  $ equal to one. To compare very short pulses
one may prefer to fix $\omega$ and $a_{0}$ so that the pulse's $\phi$ range is
of order $2\pi$ and $-\int d\phi f^{\prime2}\left(  \phi\right)  =1$. Whatever
our choice, $a_{0}$ should offer a reliable description of the peak intensity,
so $\left\vert f^{\prime}\left(  \phi\right)  ^{2}\right\vert _{\max}\sim1$.

\paragraph{Classical motion:}

Let $\pi=mu$ be the kinetic momentum of a classical electron moving in this
plane wave field, where%
\begin{equation}
u\left(  \phi\right)  =v-a_{0}f\left(  \phi\right)  +\frac{2a_{0}f\left(
\phi\right)  \cdot v-a_{0}^{2}f^{2}\left(  \phi\right)  }{2k\cdot v}k.
\label{cv}%
\end{equation}
If we set $f\left(  -\infty\right)  =0$, then $v=u\left(  -\infty\right)  $ is
indeed the velocity of the particle before meeting the wave. As opposed to
unipolar pulses that permanently accelerate the particle \cite{unip}, for the
usual whole-cycle pulses, $f\left(  \infty\right)  =0$ and $u\left(
\infty\right)  =u\left(  -\infty\right)  $, as long as RR is neglected.

\paragraph{Scattering probability:}

In the Furry picture, the total NLCS probability, averaged over the initial
electron's spin and summed over all possible spin/momentum states of the final
particles, is:%
\begin{align}
P  &  =\frac{-\alpha m^{2}}{4\pi^{2}\omega^{2}p^{-}}\int_{0}^{p^{-}}%
\frac{dk^{\prime-}}{k^{\prime-}\left(  p^{-}-k^{\prime-}\right)  }\int_{%
\mathbb{R}
^{2}}d\mathbf{k}^{\prime\perp}\nonumber\\
&  \times\int_{%
\mathbb{R}
^{2}}d\phi d\phi^{\prime}\left[  1-a_{0}^{2}g\left(  \tfrac{k^{\prime-}}%
{p^{-}}\right)  \theta^{2}\left\langle f^{\prime}\right\rangle ^{2}\right]
e^{\frac{ik^{\prime}\cdot\left\langle \pi\right\rangle \theta}{\omega\left(
p^{-}-k^{\prime-}\right)  }}, \label{prob0}%
\end{align}
where $\theta=\phi^{\prime}-\phi$,$~g\left(  \zeta\right)  =\frac{1}{2}%
+\frac{\zeta^{2}}{4\left(  1-\zeta\right)  }$ and\ we denoted the moving
average of a function $F$ by $\left\langle F\right\rangle =\theta^{-1}%
\int_{\phi}^{\phi\prime}F\left(  \xi\right)  d\xi$. The formal expression
(\ref{prob0}) was obtained from formula C1 in \cite{unip}, as follows. Instead
of the expressions C3, we used the unregularized integrals $B_{\nu}=\int_{%
\mathbb{R}
}\tilde{f}_{\nu}e^{-i\Phi}d\phi$, with $\tilde{f}_{\nu}\in\left(
1,f_{1},f_{2},\mathbf{f}^{2}\right)  $. Explicitation of all $B_{\nu}$ in the
quadratic form C2 led to the inner double integral in (\ref{prob0}), by
writing $f\left(  \phi^{\prime}\right)  -f\left(  \phi\right)  $ as
$\left\langle f^{\prime}\right\rangle \theta$. Then, the $s$ and $\bar{p}$
integrals were performed, eliminating the four dimensional delta function and
imposing the lightfront conservation laws $\bar{p}^{\nu}=p^{\nu}-k^{\nu}$,
$\nu\in\left\{  1,2,-\right\}  $. In the exponent the average of the classical
momentum $\left\langle \pi\right\rangle $ was identified. A change of variable
from $k^{3}$ to $k^{-}$ led to the final result. See also \cite{boca1, SHK,
SK}. If we want to compute them first, all $B_{\nu}$ in the generic case, or
at least $B_{0}$, need to be regulated, damping the oscillations of the
integrand with a convergence factor such as $e^{-\varepsilon\phi^{2}}$,
$\varepsilon\searrow0$, that can be discarded after a partial integration
restricts $B_{\nu}$ to the length of the pulse \cite{boca1, unip}.

At first glance, in writing (\ref{prob0}) we have added to the numerical
complexity, constructing a double integral out of simple ones. But, in fact,
by a change of quadrature order, the analytical integration over
$\mathbf{k}^{\prime\perp}$, and $k^{\prime-}$ leaves us with only two easy
integrals, instead of the initial four. In addition, we get rid of the rapid
oscillations encountered when computing $B_{\nu}$. Expressing $k^{\prime}%
\cdot\left\langle \pi\right\rangle $ in lightfront coordinates, we notice the
integral over $\mathbf{k}^{\prime\perp}$ is Gaussian, if regulated by
replacing in the exponent the factor $\theta$ by $\theta+i\varepsilon$, then
taking $\varepsilon\searrow0$. The previous damping factor is now superfluous.
Introducing the invariant number $b_{0}=\frac{2k\cdot v}{m}$, the result is%
\begin{equation}
P=\frac{-i\alpha}{\pi b_{0}}\int_{%
\mathbb{R}
^{2}}d\phi d\phi^{\prime}\int_{0}^{1}d\zeta\frac{1-a_{0}^{2}g\left(
\zeta\right)  \left\langle f^{\prime}\right\rangle ^{2}\theta^{2}}%
{\theta+i\varepsilon}e^{\tfrac{i\zeta\mu\theta}{\left(  1-\zeta\right)  b_{0}%
}},
\end{equation}
where
\begin{equation}
\mu=1+a_{0}^{2}\left(  \left\langle f\right\rangle ^{2}-\left\langle
f^{2}\right\rangle \right)  \geq1.
\end{equation}
We recognize the effective mass $M=\sqrt{\left\langle \pi\right\rangle ^{2}}$,
first introduced by Kibble, that appears in the Volkov propagator
\cite{kibble} and the Wigner function \cite{wig1, wig2}:%
\begin{equation}
\mu=\left\langle u\right\rangle ^{2}=\frac{M^{2}}{m^{2}}.
\end{equation}
A Lorentz and gauge invariant, $M$ depends on the averaging interval. For a
whole-cycle finite pulse, the mass shift $M-m$ vanishes when $\theta
\rightarrow\infty$. We now use the relation $\left(  \theta+i\varepsilon
\right)  ^{-1}=p.v.\theta^{-1}-i\pi\delta\left(  \theta\right)  $ and the fact
that $\mu$ is an even function of $\theta$,~so the result is indeed real.
Changing variables from $\phi$, $\phi^{\prime}$ to $\theta$ and $\sigma
=\frac{\phi+\phi^{\prime}}{2}$, and noticing that $\pi=\int_{%
\mathbb{R}
}dx\frac{\sin x}{x}$, we get
\begin{align}
P  &  =-\frac{2\alpha}{\pi b_{0}}\int_{%
\mathbb{R}
}d\sigma\int_{0}^{\infty}d\theta\int_{0}^{1}d\zeta\nonumber\\
&  \times\left[  \frac{\partial\ln\mu}{\partial\theta}+a_{0}^{2}\left\langle
f^{\prime}\right\rangle ^{2}\theta g\left(  \zeta\right)  \right]  \sin
\tfrac{\zeta\mu\theta}{\left(  1-\zeta\right)  b_{0}}.
\end{align}
A new analytical integration leads to:%
\begin{align}
P  &  =-\frac{2\alpha}{\pi b_{0}}\int_{%
\mathbb{R}
}d\sigma\int_{0}^{\infty}d\theta\nonumber\\
&  \times\left[  \frac{\partial\ln\mu}{\partial\theta}\mathcal{J}_{1}\left(
\frac{\mu\theta}{b_{0}}\right)  +a_{0}^{2}\left\langle f^{\prime}\right\rangle
^{2}\theta\mathcal{J}_{2}\left(  \frac{\mu\theta}{b_{0}}\right)  \right]  ,
\label{ppp}%
\end{align}
where, in terms of trigonometric integrals,
\begin{gather*}
\mathcal{J}_{1}\left(  x\right)  =-xA^{\prime}\left(  x\right)  ,~\mathcal{J}%
_{2}\left(  x\right)  =\frac{1}{8}\left[  2+x-x^{2}A\left(  x\right)  \right]
,\\
A\left(  x\right)  =\sin x\operatorname*{ci}\left(  x\right)  -\cos
x\operatorname*{si}\left(  x\right)  ,~\\
A^{\prime}\left(  x\right)  =\cos x\operatorname*{ci}\left(  x\right)  +\sin
x\operatorname*{si}\left(  x\right)  .
\end{gather*}
Notice that the Lorentz invariant (\ref{ppp}) depends on $b_{0}$, and
4-products of the values of the function $a_{0}f$, but not on $a_{0}f\cdot p$.
Interestingly, we can rewrite the probability in terms of the classical
velocity (\ref{cv}) as function of the proper time, eliminating all reference
to the driving field and proving gauge invariance. Could the result be
generalized to an arbitrary, not necessarily plane, wave? It is hard to
answer, because of the many very different trajectories allowed inside a
general field. The attractiveness of a plane wave derives from the simplicity
of the law of motion it entails. The electron's motion always looks the same,
regardless of the initial position. A quantum computation for a general field
would require the use of a wavepacket with some initial average position and
momentum that only in the limit relates to a particular classical motion.

\paragraph{Radiated energy-momentum:}

The same procedure can be applied to compute the expectation value of the
emitted photon's momentum,%
\begin{align}
\left\langle k^{\prime\nu}\right\rangle  &  =-\frac{\alpha m^{2}}{4\pi
^{2}\omega^{2}p^{-}}\int_{0}^{p^{-}}\frac{dk^{\prime-}}{k^{\prime-}\left(
p^{-}-k^{\prime-}\right)  }\int_{%
\mathbb{R}
^{2}}d\mathbf{k}^{\prime\perp}k^{\prime\nu}\nonumber\\
&  \times\int_{%
\mathbb{R}
^{2}}d\phi d\phi^{\prime}e^{\frac{ik^{\prime}\cdot\left\langle \pi
\right\rangle \theta}{\omega\left(  p^{-}-k^{\prime-}\right)  }}\left[
1-a_{0}^{2}\left\langle f^{\prime}\right\rangle ^{2}\theta^{2}g\left(
\tfrac{k^{\prime-}}{p^{-}}\right)  \right]  . \label{mom}%
\end{align}
Following the same method as for the probability, we are left with the
manifestly covariant double integral%
\begin{equation}
\left\langle k^{\prime\nu}\right\rangle =\frac{2\alpha}{\pi}\int_{%
\mathbb{R}
}d\sigma\int_{0}^{\infty}d\theta\left(  F\left\langle \pi^{\nu}\right\rangle
+Gk^{\nu}\right)  , \label{mom2}%
\end{equation}
where%
\begin{gather}
F=\frac{\mathcal{J}_{3}\left(  \frac{\mu\theta}{b_{0}}\right)  -\mathcal{J}%
_{3}\left(  \frac{\theta}{b_{0}}\right)  }{b_{0}\theta}-a_{0}^{2}\left\langle
f^{\prime}\right\rangle ^{2}\frac{\theta}{b_{0}}\mathcal{J}_{4}\left(
\frac{\mu\theta}{b_{0}}\right)  ,\label{Kaa}\\
G=\frac{1}{b_{0}^{2}}\left(  \frac{2-\mu}{\mu}\frac{\partial\mu}%
{\partial\theta}-2\frac{\mu-1}{\theta}\right)  \mathcal{J}_{3}\left(
\frac{\mu\theta}{b_{0}}\right)  -\frac{a_{0}^{2}}{b_{0}}\left\langle
f^{\prime}\right\rangle ^{2}\mathcal{J}_{5}\left(  \frac{\mu\theta}{b_{0}%
}\right)  ,
\end{gather}
and the new functions used are%
\begin{gather*}
\mathcal{J}_{3}\left(  x\right)  =\tfrac{x^{2}}{2}A\left(  x\right)
-xA^{\prime}\left(  x\right)  -\tfrac{x}{2},\\
\mathcal{J}_{4}\left(  x\right)  =\tfrac{1}{24}\left[  \left(  6x-x^{3}%
\right)  A\left(  x\right)  \right]  ^{\prime}+\tfrac{x}{12},\\
\mathcal{J}_{5}\left(  x\right)  =\tfrac{1}{24}\left[  \left(  x^{4}%
-6x^{2}\right)  A\left(  x\right)  \right]  ^{\prime}-\tfrac{x^{2}}{8}%
+\tfrac{1}{3}.
\end{gather*}
Notice that $\left\langle k^{\prime\nu}\right\rangle $ was computed with a
distribution normalized not to unity, but to $P$. Therefore it is not the
average value of the momentum of \textit{one} photon, assuming an emission has
taken place, but the expected outcome per incident electron, counting both
NLCS and ellastic scattering events. To compute the former, conditional
probabilities are needed. This amounts to dividing the expectation value by
$P$. The standard deviation of $k^{\prime\nu}$ or any of its higher moments
can similarly be computed.

\paragraph{Periodic and adiabatic limits}

Even though derived for a pulse, our formulae can provide the emission
probability and radiated energy-momentum per cycle in the case of an infinite,
periodic plane wave, described by $\tilde{f}\left(  \phi\right)  =\tilde
{f}\left(  \phi+T\right)  $. They can be obtained by truncating the pulse to a
finite number of cycles $N$, and considering the quickly reached
$N\rightarrow\infty$ limit of $P/N$ and $\left\langle k^{\prime\nu
}\right\rangle /N$. The results, that we shall denote by $\tilde{P}\left(
a_{0}\right)  $ and $\left\langle \tilde{k}^{\prime\nu}\right\rangle \left(
a_{0}\right)  $, are given by (\ref{ppp}) and (\ref{mom2}) with the $\sigma$
integral restricted to one period. A circularly polarized monochromatic wave
$\tilde{f}\left(  \phi\right)  =\left(  0,\cos\phi,\sin\phi,0\right)  $
provides a simple example, as $\mu=1+a_{0}^{2}\left(  1-\operatorname{sinc}%
{}^{2}\frac{\theta}{2}\right)  $ is independent of $\sigma$, so $\int
d\sigma\rightarrow2\pi$. Consider now the modulated wave,%

\begin{equation}
f\left(  \phi\right)  =g\left(  \frac{\phi}{\tau}\right)  \tilde{f}\left(
\phi\right)  , \label{modulat}%
\end{equation}
where $\tilde{f}\left(  \phi\right)  $ is a periodic function, e.g.
monochromatic, the envelope $g\left(  x\right)  $ is a smooth function
vanishing at infinity and $\tau$ controls the pulse length. Provided $\tau\gg
T$, (\ref{ppp}) and (\ref{mom2}) are practically proportional to $\tau$, being
well approximated by%

\begin{align}
P  &  \simeq\frac{\tau}{T}\int_{%
\mathbb{R}
}dx\tilde{P}\left[  a_{0}g\left(  x\right)  \right]  ,\label{adiap}\\
\left\langle k^{\prime\nu}\right\rangle  &  \simeq\frac{\tau}{T}\int_{%
\mathbb{R}
}dx\left\langle \tilde{k}^{\prime\nu}\right\rangle \left[  a_{0}g\left(
x\right)  \right]  . \label{adiam}%
\end{align}
\newline

Of particular interest are the limiting cases of small $a_{0}$ or $b_{0}$.
With today's technology, $\omega$ is around the order of $1-10^{4}$ $eV$ and
only for the optical range a large $a_{0}$ is possible. If, for instance,
$\omega=3~eV$, an electron energy of at least $20$~$GeV$ is needed for $b_{0}$
to reach unity.

\paragraph{Perturbative limit}

The results can be expressed as integrals containing the field's Fourier
transform and the Klein-Nishima probability of linear Compton \cite{unip}.
However, to emphasize the role of $b_{0}$, we prefer to work with a quadratic
function related to the potential's auto-correlation,%
\begin{equation}
\mathcal{E}\left(  \theta\right)  =-a_{0}^{2}\int_{%
\mathbb{R}
}d\sigma\left\langle f^{\prime}\right\rangle ^{2}. \label{eee}%
\end{equation}
The weak field limits ($a_{0}\ll1$) of (\ref{ppp}) and (\ref{mom2}) are:%
\begin{gather}
P_{p}=\frac{2\alpha}{\pi}\int_{0}^{\infty}d\theta\mathcal{P}\left(
\tfrac{\theta}{b_{0}}\right)  \mathcal{E}\left(  \theta\right)  ,\label{pclas}%
\\
\left\langle k^{\prime\nu}\right\rangle _{p}=\frac{2\alpha}{\pi}\int
_{0}^{\infty}d\theta\left[  p^{\nu}\mathcal{F}\left(  \tfrac{\theta}{b_{0}%
}\right)  +k^{\nu}\mathcal{G}\left(  \tfrac{\theta}{b_{0}}\right)  \right]
\mathcal{E}\left(  \theta\right)  ,
\end{gather}
where $\mathcal{P}$, $\mathcal{F}$, and $\mathcal{G}$ are the universal,
positive functions,%
\begin{align*}
\mathcal{P}\left(  x\right)   &  =x^{2}\int_{x}^{\infty}\tfrac{y-2x}{y^{3}%
}\mathcal{J}_{1}\left(  y\right)  dy+x\mathcal{J}_{2}\left(  x\right)  ,\\
\mathcal{F}\left(  x\right)   &  =x^{2}\int_{x}^{\infty}\tfrac{y-2x}{y^{3}%
}\mathcal{J}_{3}\left(  y\right)  dy+x\mathcal{J}_{4}\left(  x\right)  ,\\
\mathcal{G}\left(  x\right)   &  =x^{2}\int_{x}^{\infty}\tfrac{4x-3y}{y^{3}%
}\mathcal{J}_{3}\left(  y\right)  dy+\mathcal{J}_{5}\left(  x\right)  .
\end{align*}
Since $\mathcal{P}\left(  x\right)  $ is increasing for $x>0$, \ref{pclas}
decreases with $b_{0}$, having the classical upper bound%

\[
P_{p,\ cl}=\frac{2\alpha}{3\pi}\int_{0}^{\infty}d\theta\mathcal{E}\left(
\theta\right)  .
\]

\paragraph{Classical limit:}

This is expected when $b_{0}\ll1$, since $b_{0}$ is proportional to $\hbar$, a
fact obscured by the use of natural units. That is, the laser photon energies
are much smaller than\ $m$ in the electron's rest frame. Now the major
contribution to the integral (\ref{mom2}) comes from small $\theta$. By Taylor
expansions, such as%
\begin{equation}
\mu=1+a_{0}^{2}\left[  \frac{-f^{\prime}\left(  \sigma\right)  ^{2}}{12}%
\theta^{2}+\mathcal{O}\left(  \theta^{4}\right)  \right]  , \label{masss}%
\end{equation}
one obtains the approximation%
\begin{equation}
\left\langle k^{\prime\nu}\right\rangle \simeq-\frac{\alpha}{3}b_{0}a_{0}%
^{2}\int_{%
\mathbb{R}
}d\sigma\pi^{\nu}\left(  \sigma\right)  f^{\prime}\left(  \sigma\right)
^{2}\mathcal{C}\left[  -\tfrac{b_{0}^{2}a_{0}^{2}f^{\prime}\left(
\sigma\right)  ^{2}}{12}\right]  , \label{class0}%
\end{equation}
where%

\begin{equation}
\mathcal{C}\left(  x\right)  =\frac{1}{\pi}\int_{0}^{\infty}\left[
6\mathcal{J}_{4}\left(  y+xy^{3}\right)  -\mathcal{J}_{3}\left(
y+xy^{3}\right)  \right]  ydy. \label{corr}%
\end{equation}
No classical equivalent is found for the second term in the integrand of
(\ref{mom2}), negligible in this limit. An equivalent form of (\ref{class0})
can be written by expressing everything in terms of a classical particle's
velocity as function of proper time,%
\begin{equation}
\left\langle k^{\prime\nu}\right\rangle \simeq-\frac{1}{6}\frac{e^{2}}{\pi
}\int_{%
\mathbb{R}
}d\tau u^{\nu}\left(  \tau\right)  \dot{u}\left(  \tau\right)  ^{2}%
\mathcal{C}\left[  -\tfrac{\dot{u}\left(  \tau\right)  ^{2}}{3m^{2}}\right]  .
\label{class1}%
\end{equation}
When not only $b_{0}$, but also $b_{0}a_{0}$ is very small, equation
(\ref{class1}) further reduces to:%
\begin{equation}
\left\langle k^{\prime\nu}\right\rangle _{cl}=-\frac{1}{6}\frac{e^{2}}{\pi
}\int_{%
\mathbb{R}
}d\tau u^{\nu}\left(  \tau\right)  \dot{u}\left(  \tau\right)  ^{2}.
\label{class2}%
\end{equation}

The same result arises directly from classical electrodynamics, if one
neglects radiation reaction. For $\nu=0$, (\ref{class2}) is just the time
integrated Larmor's formula.%

\begin{figure}
[ptb]
\begin{center}
\includegraphics[
height=2.2978in,
width=3.3589in
]%
{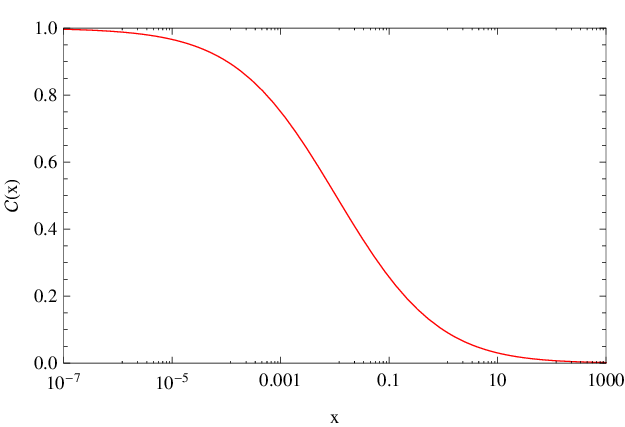}%
\caption{(color online) The function describing the strong field correction to
Larmor's formula}%
\label{corectie}%
\end{center}
\end{figure}
The function (\ref{corr}) is decreasing for $x>0$. As shown in Fig.
\ref{corectie}, it departs very quickly from its upper bound $\mathcal{C}%
\left[  0\right]  =1$. The deviation is noticeable even for $x=10^{-5}$. It
follows that a small $b_{0}$ doesn't necessarily make classical Thompson
scattering a good model. For strong fields, much less power is radiated than
Larmor predicts. In CED, the spectral and angular distribution is also a
double integral, but an interesting cancelling of interference terms leaves a
single one after the frequency is integrated away. The transfer of energy and
momentum to the field is well defined at each moment in time, there are no
quantum uncertainties. Interestingly, this form of decoherence is also shown
by the better approximation (\ref{class1}), that looks misleadingly classical,
though the argument of the correction $\mathcal{C}$ is in fact proportional to
$\hbar^{2}$. Heuristically, $b_{0}$ is a coherence lighfront time $\theta$
scale. When small enough, it allows for a time-incoherent model of emission,
but at high intensities the mass shift implied by (\ref{masss}) cannot be
neglected in this coherence interval, hence the aforementioned correction.
This happens when the peak electric field in the electron rest frame
approaches the Schwinger critical value $m^{2}/e$. Formula (\ref{class1})
could provide a general improvement to Larmor's, valid in an arbitrary driving
field, whose frequencies in the rest frame of the electron are similarly low
compared to $m$, so the emission can be viewed as incoherent in time, hence
the product of the classical motion of a charged particle. This should not be
confused to the second term in the radiated energy's expansion in powers of
$\hbar$, at fixed $a_{0}$, that is non-local in time, as discussed in
\cite{larm1, larm2}. As for the total scattering probability, by an asymptotic
expansion of the functions $\mathcal{J}_{i}\left(  x\right)  $ one gets the
limit%
\begin{equation}
P_{cl}=\frac{2\alpha}{\pi}a_{0}^{2}\int_{0}^{\infty}\frac{d\theta}{\theta^{2}%
}\int_{%
\mathbb{R}
}d\sigma\frac{\left\langle f^{2}\right\rangle -\left\langle f\right\rangle
^{2}-\frac{1}{2}\left\langle f^{\prime}\right\rangle ^{2}\theta^{2}}%
{1+a_{0}^{2}\left(  \left\langle f\right\rangle ^{2}-\left\langle
f^{2}\right\rangle \right)  }.\label{pc}%
\end{equation}
In this case no decoherence is observed, as photon emission probability is not
a classical concept.

\paragraph{Numerical results:}

We start illustrating our results using a one-cycle, linearly polarized pulse,
characterized by a Gaussian potential:%
\begin{equation}
f^{\mu}\left(  \phi\right)  =\sqrt[4]{\tfrac{2}{\pi}}\exp\left(
-x^{2}\right)  \delta_{\mu1}. \label{pul}%
\end{equation}
%

\begin{figure}
[ptb]
\begin{center}
\includegraphics[
height=2.1802in,
width=3.3641in
]%
{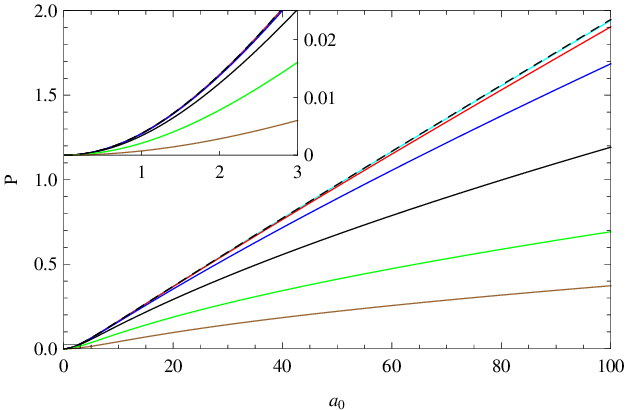}%
\caption{(color online) The NLCS probability (\ref{ppp}) is plotted as
function of $a_{0}$, for the pulse (\ref{pul}) and $b_{0}=10^{p}$,
$p=-4,-3,...,1$. The dashed line is the classical limit (\ref{pc}). The inset
details the perturbative region of small $a_{0}$.}%
\label{probota}%
\end{center}
\end{figure}

The total NLCS probability is shown in Fig.\ref{probota}. The quadratic
increase from the perturbative region, shown in detail in the upper left
corner, quickly slows down as $a_{0}$ grows past unity. The higher the
parameter $b_{0}$, the lower $P$ is. \ In general, (\ref{ppp}) boundlessly
grows with the length/intensity of the pulse. Even for one as short as
(\ref{pul}) and the experimentally attainable $a_{0}=100$, the result can
easily surpass unity. In \cite{piazza}, this possibility was noticed,
interpreted as a sign that multiphoton emission cannot then be neglected, and
a re-normalization of the whole series of n-photon NLCS probabilities was
suggested. Moreover, for a unipolar pulse, (\ref{ppp}) shows the logarithmic
IR divergence typical of Bremsstrahlung \cite{unip, greger}. These problems
can be dealt with by including a one loop self-energy diagram, that adds
nothing to (\ref{mom2}), but does contribute to the expectation value of the
final electron's momentum, even in the whole-cycle case \cite{rrqed}. For the
general theory of the cancellation between real and virtual photon IR
divergences, see \cite{FSY}.%

\begin{figure}
[ptb]
\begin{center}
\includegraphics[
height=2.1811in,
width=3.3641in
]%
{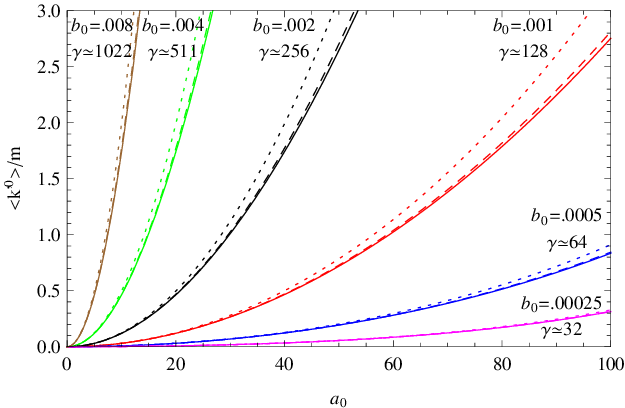}%
\caption{(color online) The expectation value of $\left\langle k^{\prime
0}\right\rangle $, as a multiple of the electron rest energy, plotted as
function of $a_{0}$, for the pulse (\ref{pul}). The dotted line is the
Thompson limit (\ref{class2}), while the dashed one is the better
approximation (\ref{class1}).}%
\label{ene}%
\end{center}
\end{figure}

To plot formula (\ref{mom2}), we need more than just the invariants $a_{0}$,
$b_{0}$. The frequency of any available laser that can reach the nonlinear
regime $a_{0}\gtrsim1$ is around the optical range. Let's assume $\omega
=1~eV$. It remains to know the incidence, that we choose head-on. While
experimentally difficult, this gives the largest $b_{0}$ for a given pulse and
electron beam. In Fig. \ref{ene} a comparison is drawn between the expectation
value of the radiated energy and its two incoherent approximations. Both
(\ref{class2}) and (\ref{class1}) overestimate (\ref{mom2}), but the latter is
a much closer match.

\begin{figure}
[ptb]
\begin{center}
\includegraphics[
height=3.3589in,
width=3.3589in
]%
{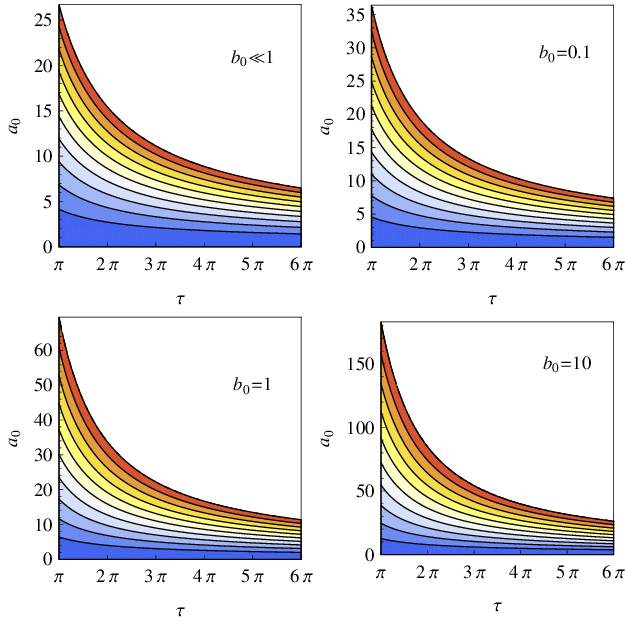}%
\caption{(color online) Contour plot of the NLCS probability (\ref{ppp}), in
terms of the duration/strength of the pulse (\ref{puls}), for linear
polarization and $\phi_{0}=0$. The contours correspond to $P=0.1,~0.2,...,1.$}%
\label{probitate}%
\end{center}
\end{figure}

\begin{figure}
[ptb]
\begin{center}
\includegraphics[
height=2.1733in,
width=3.3624in
]%
{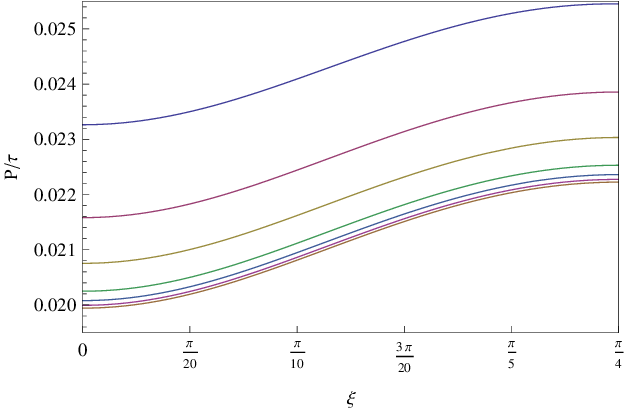}%
\caption{(color online) The probability (\ref{ppp}) divided by the length of
the pulse (\ref{puls}), as function of polarization state, for $a_{0}=5$,
$b_{0}=1$ and $\phi_{0}=0$. The curves correspond, in decreasing order, to
$\tau/\pi=1,2,4,10,20,40,100$.}%
\label{adia}%
\end{center}
\end{figure}

Let us now consider a modulated harmonic wave,%

\begin{equation}
f\left(  \phi\right)  =g\left(  \frac{\phi}{\tau}\right)  \left(  0,\cos
\xi\sin\left(  \phi-\phi_{0}\right)  ,\sin\xi\cos\left(  \phi-\phi_{0}\right)
,0\right)  \text{,} \label{puls}%
\end{equation}
where $\xi$ describes the polarization state, $\phi_{0}$ is known as
carrier-envelope phase (CEP) and $\tau$ controls the pulse length. We present
computations for the symmetric, Gaussian envelope $g\left(  x\right)
=e^{-x^{2}}$. For the carrier-envelope model to make sense, we assume $\tau$
is larger than, say, $\pi$. Fig. \ref{probitate}, shows contour plots of the
total probability (\ref{ppp}) as function of pulse peak potential/length, for
linear polarization and various values of the nonclassicality parameter
$b_{0}$. Again, we find that in experimentally realistic conditions, our
result can easily surpass unity, signalling the need for taking into account
higher order Feynman diagrams. We are interested in the region where $P$ stays
well below unity, so the model can be given credit. This region grows with
$b_{0}$, once it gets close to unity. The influence of the polarization state
and the convergence towards the adiabatic limit (\ref{adiap}), as the pulse
length increases, are shown in Fig. \ref{adia}. The CEP $\phi_{0}$ has a very
small impact on $P$ for this smooth potential.%

%

\begin{figure}
[ptb]
\begin{center}
\includegraphics[
height=2.2874in,
width=3.3598in
]%
{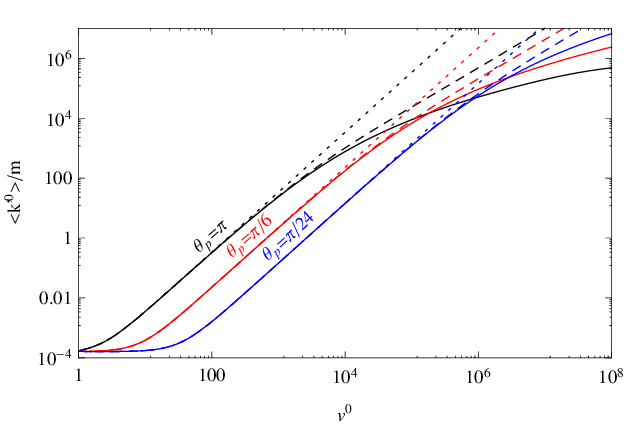}%
\caption{(color online) A plot of the expectation value $\left\langle
k^{\prime0}\right\rangle $ versus the initial electron's Lorentz factor for
the pulse (\ref{puls}) with $\tau=3\pi$, $\xi=\phi_{0}=0$ and $a_{0}=10$. The
incidence angles are $\theta_{p}=\pi,~\frac{\pi}{6},~\frac{\pi}{24}$ and
$\phi_{p}=0$. The head-on result is much larger at low energies, surpassed by
the others at very high ones. The dotted/dashed lines are the approximations
(\ref{class2}) and (\ref{class1}).}%
\label{loglog}%
\end{center}
\end{figure}

\begin{figure}
[ptb]
\begin{center}
\includegraphics[
height=2.3203in,
width=3.3589in
]%
{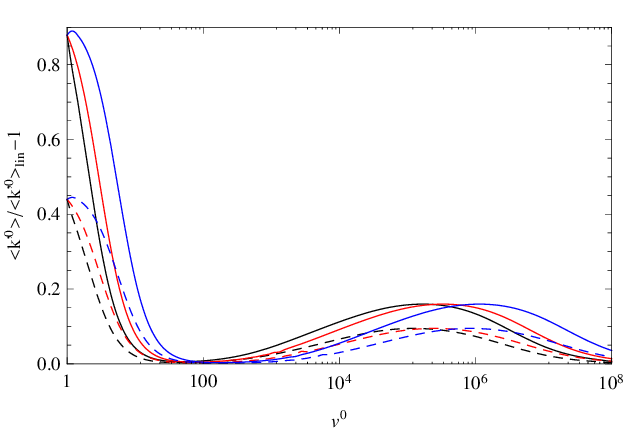}%
\caption{(color online) The relative increase in the average radiated energy
when the polarization is changed from linear to circular ($\xi=\frac{\pi}{4}$,
solid line) / elliptic ($\xi=\frac{\pi}{8}$, dashed line). The pulse is
(\ref{puls}), $\tau=3\pi$, $\phi_{0}=0$, $a_{0}=10$, $\phi_{p}=0$ and
$\theta_{p}=\pi$, $\frac{\pi}{2}$, $\frac{\pi}{4}$ (growing order at left)}%
\label{pol}%
\end{center}
\end{figure}

\begin{figure}
[ptb]
\begin{center}
\includegraphics[
height=2.2978in,
width=3.3806in
]%
{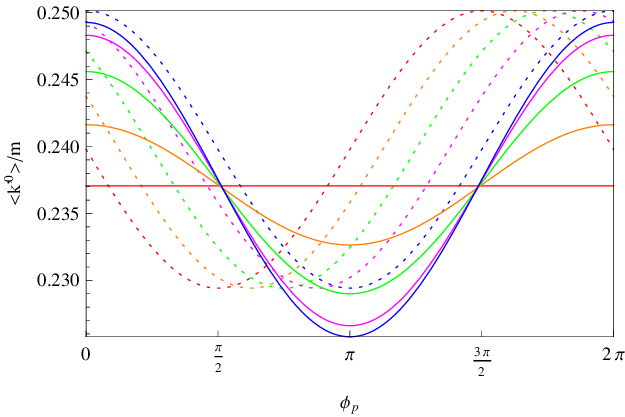}%
\caption{(color online) The radiated energy's dependence on the angle
$\phi_{p}$, for $\theta_{p}=\frac{\pi}{2}$, $\gamma=100$ and the pulse
(\ref{puls}), with $\tau=\pi$, $a_{0}=20$ and CEP $\phi_{0}=0,\frac{\pi}%
{8},\frac{\pi}{4},\frac{3\pi}{8},\frac{\pi}{2}$ (growing order at left).
Solid/dotted stand for circular/linear polarization.}%
\label{jug}%
\end{center}
\end{figure}

In order to study (\ref{mom2}), we now choose an optical frequency of $3$
$eV$, and specify the initial electron momentum, setting its Lorentz factor
$v^{0}$ and polar angles $\theta_{p}$ and $\phi_{p}$. We have already looked
at the energy $\left\langle k^{\prime0}\right\rangle $ for a one-cycle pulse,
so we consider a longer one. In Fig. \ref{loglog} a plot of the radiated
energy's dependence on the initial electron energy is shown for the pulse
(\ref{puls}) with $\tau=3\pi$, linear polarization and $a_{0}=10$. In the
classical regime, after $v^{0}$ becomes much greater than $a_{0}$, the growth
is quadratic, but at even higher energies quantum effects slow it down. For a
while, it is well described by the local in time, incoherent approximation
(\ref{class1}). Then, as $b_{0}\,$goes past unity, we reach a highly nonlocal
quantum radiation regime. It is interesting to notice that, at very high
energies, the emission becomes stronger for lower incidence angles $\theta
_{p}$ than for higher ones, because the corresponding smaller $b_{0}$ implies
weaker quantum effects. Fig. \ref{pol} shows the increase of $\left\langle
k^{\prime0}\right\rangle $ as the polarization is changed from linear towards
circular. The influence of $\phi_{p}$ on (\ref{mom2}) is due to the $f\left(
\phi\right)  \cdot v$ term in (\ref{cv}). For the example considered in Fig.
\ref{loglog}, both this dependence and the one on the CEP $\phi_{0}$ are
extremely weak. They can be noticeable for a very short pulse and small
$b_{0}$, as seen from the example in Fig. \ref{jug}. As to the vector
$\left\langle \mathbf{k}^{\prime}\right\rangle $, for large $v^{0}$, it
practically has the direction of $\mathbf{v}$ and the length $\left\langle
k^{\prime0}\right\rangle $, in the laboratory frame all emission being
concentrated inside a very narrow cone. Significant differences arise only at
low energies, where quantum effects are, however, small.%

%

%

\paragraph{Conclusions:}

We have found a way to analytically integrate all final state variables out of
the NLCS probability and expectation values. This not only allowed for the
first time their detailed numerical exploration, by saving huge computational
expense, but also revealed new insights into the structure of scattering
processes in strong fields. We shed light on the role of the effective mass
and the emission's coherence in time. Simple results were found for the
monochromatic, perturbative and classical limits. We derived a strong field
correction to Larmor's formula, arising from the mass shift. Computations
performed under realistic conditions showed its usefulness, but also found
large values for the probability, even surpassing unity. This suggests
multiple scatterings and radiative corrections need to be considered. In a
future paper, our method will be applied to these, as well as to other strong
field processes.

\paragraph{Acknowledgements}

The author thanks V. Florescu, A. Ilderton and G. Torgrimsson for useful
discussions, and ESF-RNP-SILMI for support in attending conference FILMITh,
Garching 2012

\end{document}